\documentclass[12pt]{article}
\usepackage{graphicx,amsmath}

\newlength{\dinwidth}
\newlength{\dinmargin}
\renewcommand{\vec}[1]{\boldsymbol{#1}}

\def\lapproxeq{\lower .7ex\hbox{$\;\stackrel{\textstyle                                                    
<}{\sim}\;$}}                                                    
\def\gapproxeq{\lower .7ex\hbox{$\;\stackrel{\textstyle                                                    
>}{\sim}\;$}}                                                    
\def\be{\begin{equation}}                                                    
\def\ee{\end{equation}}                                                    
\def\bea{\begin{eqnarray}}                                                    
\def\eea{\end{eqnarray}}                                                    

\def\slash#1{#1 \hskip-0.55em /}
\def\Pslash#1{#1 \hskip-0.6em /}
\def\sh{\hat s}
\def\sh2{{\hat s}^2}
\begin{document}                                                    
\titlepage                                                    
\begin{flushright}                                                    
IPPP/10/14  \\
DCPT/10/28 \\                                                    
\today \\                                                    
\end{flushright}                                                    
                                                    
\vspace*{0.5cm}                                                    
                                                    
\begin{center}                                                    
{\Large \bf A new window at the LHC:\\}
\vspace*{0.2cm}
{\Large \bf BSM signals using tagged protons}                                                                                                        
                                                    
\vspace*{1cm}                                                    
V.A. Khoze$^{a,b}$, A.D. Martin$^a$,  M.G. Ryskin$^{a,c}$ and A.G. Shuvaev$^{c}$  \\                                                    
                                                   
\vspace*{0.5cm}                                                    
$^a$ Institute for Particle Physics Phenomenology, University of Durham, DH1 3LE \\                                                   
$^b$ School of Physics \& Astronomy, University of Manchester, Manchester M13 9PL\\ 
$^c$ Petersburg Nuclear Physics Institute, Gatchina, St.~Petersburg, 188300, Russia\\           
                                                    
\end{center}                                                    
\vspace*{1cm}                                                    
                                                    
\begin{abstract}
The signature at the LHC of many Beyond the Standard Model (BSM) scenarios is events with large missing energy. If the forward outgoing protons are measured, we show that the production and decay of BSM particles in the central rapidity interval, with gaps in rapidity either side, offers certain advantages over inclusive production, to search for signals (a) with missing longitudinal 4-momentum (typical of invisible Higgs production), and (b) for new light pseudoscalar bosons. 

\end{abstract}    

\section{Introduction}

It challenging to observe new (BSM) physics in inclusive processes at the LHC, where the signal 
 could be buried under the overwhelming backgrounds.
 In the present paper we discuss the possibility to enhance the signal/background ratio using tagged outgoing forward protons. Actually, by selecting events where the momentum fraction of the outgoing proton is very close to 1, we turn the LHC into a Pomeron-Pomeron (or $\gamma\gamma$) collider, see, for instance\cite{KMRpros}. 
Of course, the energy and the luminosity are lower, but this may be compensated by the advantages which we discuss in Section 2.

We consider some feasible applications in Sections 3-6. In Section 3 we discuss `invisible' Higgs detection. Another application is the possibility of using the semi-inclusive process $pp \to p+g\phi_- g+p$ to observe the production of pseudoscalar bosons, $\phi_-$, where the $+$ signs denote the presence of large rapidity gaps. Recall that in the pure exclusive process $pp \to p+\phi_- +p$, pseudoscalar production is strongly suppressed by a ``$J_z=0,$ P-even'' selection rule \cite{KMR19}. However, in Section 4, we show that this is not the case for the semi-inclusive process, see also \cite{KMRextend}.
 In this Section we consider the cross sections for both pseudoscalar $\phi_-$ and scalar $\phi_+$ semi-inclusive production. We  apply these results, first, in Section 5, to the semi-inclusive production of the neutral bosons which occur in the Higgs sector of supersymmetric models, where, for example, we consider the production of the pseudoscalar $A$ and the scalars $h,~H$ of the MSSM; and then, second, in Section 6, we study the possible signals for the relatively light pseudo-Goldstone bosons which may result from the spontaneous breaking of some new symmetry at very high scales, which are a typical feature of many BSM scenarios, see,
for example, \cite{dg}~-~\cite{cc2}.

\section{Advantages of semi-inclusive processes}

The experimental signatures of many Beyond the Standard Model (BSM) scenarios are characterised by events with sizeable missing energy and momentum, see, for example, \cite{miss}.
 For example, in SUSY we may produce a pair of massive supersymmetric particles (gluinos or squarks) at the LHC, each of which decay into known particles together with the lightest supersymmetric particle which is stable and escapes detection. Another missing-energy BSM scenario is Large Extra Dimensions (LED) in which interactions at the LHC may produce (Kaluza-Klein) gravitons, $G$, which disappear into bulk space \cite{extrad}. Yet another example is the production of a Higgs boson which decays into a 4th generation heavy neutrino pair, for instance,
\cite{PF}~-~\cite{BKMR} or a heavy photon pair \cite{HMN}, or other neutral particles which are not observable in the detectors, see, for example \cite{DG,AM}. There are other examples of a very specific experimental signature for the
Higgs boson decay, such as diphoton plus missing energy, considered in \cite{MMP}.
 Of course, unexpected scenarios could well give events with missing energy and momentum. 

\begin{figure} [t]
\begin{center}
\includegraphics[height=6cm]{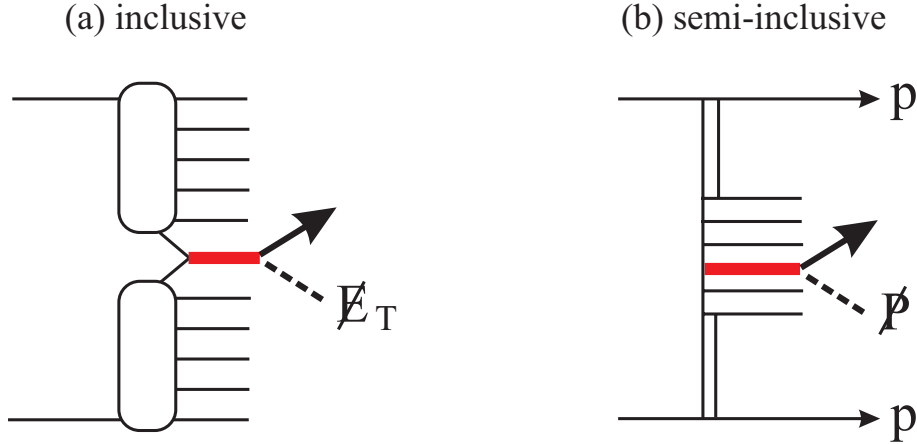}
\caption{\sf The inclusive and semi-inclusive production of a BSM particle which subsequently decays with large missing transverse energy $\slash{E}_T$ or missing 4-momentum, $\Pslash{P}$. The figure shows decays of the BSM object where some of the decay products are visible in the central detector. The method applies equally to totally invisible decays, as discussed in our illustrative example.}
\label{fig:f1}
\end{center}
\end{figure}

The inclusive signal (Fig. \ref{fig:f1}(a)) of such BSM missing-transverse-energy events have sizeable SM backgrounds, particularly at lower values of $\slash{E}_T$ coming, for example, from $(Z\to \nu \bar{\nu})+X$ or $c\bar{c}+X$ production where the charm, $c$, quark decays semi-leptonically. The neutrinos escape undetected. Here we discuss a much cleaner missing-energy signal at the LHC, provided the new objects are not too heavy, see also \cite{KMRpros,DKMOR,royon}.
We consider the production of a massive system in some central rapidity interval, with large rapidity gaps on either side, with the forward outgoing protons measured at some distance 
 far from the interaction point\footnote{Projects to install the proton detectors at 220~m and 420~m from the
interaction points are now under review in ATLAS and
CMS~\cite{CMS-Totem}~-~\cite{AFP}.}. 
 Actually, in this case, we can measure the missing 4-momentum, $\Pslash{P}$. The process is sketched in Fig. \ref{fig:f1}(b). Thus, Fig. \ref{fig:f1} shows the presence of BSM particles via decays either with missing transverse energy, $\slash{E}_T$, or with missing 4-momentum, $\Pslash{P}$.

At first sight it might appear strange to convert the LHC into a lower energy, lower luminosity `Pomeron-Pomeron' collider (or $\gamma\gamma$ collider). However there are advantages:
\begin{itemize}
\item We have almost $4\pi$ geometry of detection of the centrally produced system\footnote{Emission into the beam pipe (large $|\eta|$) is 
strongly suppressed by the large momentum fraction $x_L$ of the forward protons. The energy available for particles in the centrally produced system ($E_{\rm max}=E_{\rm beam}(1-x_L)$) would be too small.}. Of course, the
central detector acceptance in pseudorapidity ($|\eta|\lapproxeq 2.5$ for
tracking and $|\eta|\lapproxeq 5$ for calorimetry information)
limits, in turn, the mass of the centrally produced system that can be {\it completely} seen in the central detector. The hermiticity of the detector means we should be able, in principle, to accurately measure the missing 4-momentum, $\Pslash{P}$, and not just $\slash{E}_T$. In this way, we may see a peak in the missing-mass distribution\footnote{At high LHC luminosity we have a so-called overlap
(pile-up) background caused by the coincidence between our
semi-inclusive interaction and a simultaneous interaction of another pair of protons in the same bunch crossing.
Using precise time-of-flight proton detectors may allow
the reconstruction of the  vertex position corresponding  to an inelastic  (Pomeron-Pomeron) central
interaction. This would provide a strong reduction of such an accidental coincidence
background, see \cite{FP420,bonnet}. However, only tracks of charged particle may be assigned to a vertex. The calorimeter information on neutral particles from pile-up events will not resolve the vertex, and will broaden but not displace the missing-mass peak calculated from $\Pslash{P}$. Therefore, only at the lower (pile-up free) luminosities may we see a distinct peak in the missing-mass distribution. 
In summary, the missing 4-momentum is determined from $P_{\rm incoming}-P_{{\rm forward}~p'{\rm s}}-\sum_i P_i$, where the last term allows for the possibility of accompanying particles $i$ in the central detector. This semi-inclusive process enhances the event rate by up to two orders of magnitude, as compared to exclusive production, but simultaneously degrades the missing-mass resolution, particularly in the presence of pile-up at high LHC luminosity.}, indicating the presence of a new BSM object. Note, however, the peak in a semi-inclusive process will not be so narrow as in the pure exclusive case. 
\item There is a lower SM background. In inclusive missing $E_T$ events the background originates mainly from $Z\to \nu\bar{\nu}$ and $W\to l\nu$. Their rate is suppressed for the semi-inclusive as compared to the inclusive case, since, unlike $pp$ collisions, in Pomeron-Pomeron interactions there are no valence-quark initiated subprocesses.
\item Moreover, by selecting semi-inclusive events with large rapidity gaps via the
tagged forward protons, we eliminate the underlying-event background caused by the multiple interactions of the incoming protons\footnote{We assume here that the so-called overlap
(pile-up) background could be reduced to a tolerable level. We comment on this later.}.
\item There is the possibility to scan the energy of the centrally produced system, $\sqrt{s}_{\rm central}$, and to measure the position of the threshold for the production, and hence the mass, of the new object.
\item In comparison with central exclusive production (CEP)\footnote{For example, Higgs production via $pp\to p+H+p$,
see \cite{DKMOR,KMRcan}.}
 the cross section is much larger due to the almost total absence of the Sudakov factor, $T={\rm exp}(-S)$. In fact, only the small part of $S$, due to emission in the beam pipe direction (which may escape detection in the calorimeter), should be kept. Moreover there is very little depletion of the signal due to enhanced soft rescattering (which could have populated the rapidity gaps); the survival factor $S_{\rm enh}\simeq 1$ for $M\sim 0.1\sqrt{s}$ \cite{KMRnns2}.
\item Besides having an effective `Pomeron-Pomeron' collider, there is also the possibility to study $\gamma\gamma$ central production,
see, for instance, \cite{KMRpros},\cite{KP1}$-$\cite{DKP}. 
If we are able to select events with outgoing protons with transverse momenta $\lapproxeq$ 100 MeV or $\gapproxeq$ 300 MeV, we may distinguish between $\gamma\gamma$ and `Pomeron-Pomeron' collisions, see, for example, \cite{cww}. 
\end{itemize}

If the new object is heavy ($\gapproxeq 400$ GeV), the SM background is not a danger and it is better to search for such an object in the conventional inclusive process, where the whole initial $pp$-collision 
 energy can be used. However, for objects with mass $\lapproxeq 200-300$ GeV, the lower background, together with the possibility to scan the energy, means that the semi-inclusive process offers a better opportunity for detection. Moreover, by observing the forward outgoing protons and measuring the whole missing 4-momentum, $\Pslash{P}$, we can study events with ${\it low}$ missing $E_T$, but large missing $E_L$, and have the possibility of observing peaks in the missing mass distribution.
The proposed approach will provide a valuable addition to the existing wide programme
of BSM studies in CEP processes at the LHC in the forward proton mode, see \cite{KMRpros,KMRextend},\cite{HKRSTW}$-$\cite{ismd}.

\section{Signal for invisible Higgs bosons}

At present, various BSM scenarios are under consideration. Note, that
  the current limits on the mass of the SUSY particles (the gluinos and squarks) are sufficiently large to justify using the inclusive signal. For a specific example of a BSM scenario where the semi-inclusive signal is favoured, we consider the production of the so-called invisible Higgs boson, see \cite{BKMR}.
 The expected cross section\footnote{The cross section was calculated following the prescription for central inelastic ($C$-$inel$) production in Ref. \cite{KMRpros}, but with a better account of the kinematics of the central detector.} at the LHC ($\sqrt{s}=14$ TeV) for Higgs production and decay into a 4th generation neutrino pair is as large as 350 (100) fb, if the mass is $M_H=150$ (250) GeV, after we allow for accompanying particle emission in the $|\eta|<2.5$ rapidity interval. Recall that the $E_T$ of such a boson will be rather low $(E_T \ll M_H)$, so conventional detection based on large $\slash{E}_T$ is practically impossible. The SM backgrounds for semi-inclusive central production are suppressed and can be controlled. The main background is the production of $b\bar{b}$ and $c\bar{c}$ pairs with subsequent semi-leptonic decays\footnote{Unlike the CEP process, these QCD backgrounds are not suppressed by a $J_z=0$ selection rule.}. This background is larger than the expected signal by about a factor of 300, but note that only the part of this background under the Higgs missing mass peak is relevant. Moreover, it can be rejected by observing the accompanying charged leptons, and also by observing the vertices of $B$, and possibly $D$, meson decay. Recall that if we change the Pomeron-Pomeron energy, $\sqrt{s}_{\rm central}$, we should be able to observe the threshold behaviour of the signal, which would confirm the presence of the invisible object indicated by the peak in the missing-mass distribution obtained from $\Pslash{P}$ \footnote{Here, the cross section is large enough to work at lower (pile-up free) LHC luminosities, and so avoid degradation of a missing-mass peak.}.  The possibility of detecting an invisible Higgs has been taken just an illustrative example of the use of the semi-inclusive signal. The points made above apply quite generally to semi-inclusive signatures.

\section{Semi-inclusive pseudoscalar production}

In many BSM scenarios it is natural to have a relatively light pseudoscalar boson $\phi_-$, which may result from the spontaneous breaking of some extra symmetry at high scale. As a rule such a pseudo-Goldstone boson is dominantly coupled to very heavy fermions (of this extra symmetry), and therefore the production cross section is expected to be quite small in $pp$ collisions (that is, in light quark collisions). The best possibility to create $\phi_-$ is via gluon-gluon fusion mediated by the anomaly in the heavy fermion loop. This is analogous to Higgs boson production via the top-quark triangle.  The process $gg \to \phi_-$ may be described by the effective point-like Lagrangian \cite{FW,SVZ}
\be
{\cal L}~=~\frac{C}{2}\phi_-{\rm Tr}(G_{\mu\nu}^\alpha G^{\mu\nu}_\alpha).
\label{eq:ggV}
\ee

\begin{figure} [t]
\begin{center}
\includegraphics[height=6cm]{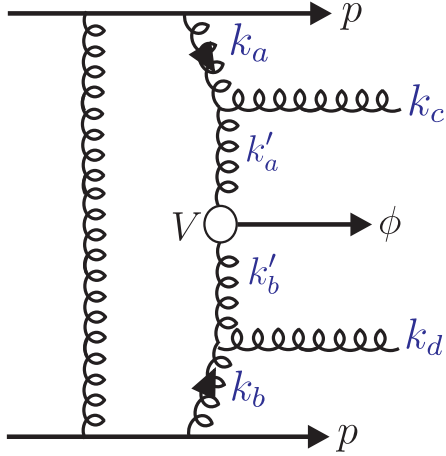}
\caption{\sf Schematic diagram for pseudoscalar ($\phi_-$) and scalar ($\phi_+$) boson production by the semi-inclusive process $pp \to p+g\phi g+p$. The gluons of the $gg^{PP} \to g\phi g$ subprocess are indicated by their 4-momenta $k_i$. }
\label{fig:a3}
\end{center}
\end{figure}
We shall see that semi-inclusive processes provide an attractive possibility to search for new pseudoscalars, $\phi_-$. The reasons are (a) that we deal with initial gluon-gluon collisions, and (b) we have much lower QCD and multiple interaction backgrounds. At first sight, we would expect a strong suppression of pseudoscalar $\phi_-$ production. The subprocess may be written as $gg^{PP} \to \phi_-$, where the superscript $PP$ indicates that the initial active gluons each come from a colour-singlet perturbative Pomeron; hence their polarisations are correlated in such a way that the incoming state has $J_z=0$ and even P \cite{KMR19}. Indeed, for the purely exclusive
 process, $pp \to p+\phi +p$, this ``$J_z=0,$ P-even'' selection rule effectively filters out the $\phi_-$ boson, leaving just
 the $\phi_+$ signal.
 However, in the semi-inclusive process, $pp \to p+g\phi g+p$, the pseudoscalar $\phi_-$ is much less suppressed  \cite{KMRextend}.
 The process is sketched in Fig.\ref{fig:a3}. The subprocess $gg^{PP} \to g\phi g$ is the lowest-$\alpha_s$-order of semi-inclusive $\phi$ production. We are unable to emit just one gluon since the initial $gg^{PP}$ state is colourless. 

The cross sections for scalar and pseudoscalar boson production, via the subprocess $gg^{PP} \to g\phi g$, are derived in the Appendix. The result is
\be
d\hat{\sigma} (gg \to g_c \phi g_d)~=~N~\frac{dk_{ct}^2}{k^2_{ct}}~\frac{dx_c}{x_c}~\frac{dk_{dt}^2}{k^2_{dt}}~\frac{dx_d}{x_d}~|{\cal M}(\phi)|^2
\label{eq:sighat}
\ee
where
\be
{\cal M}(\phi_\pm)~=~1\pm\frac{m^4_{\phi_\pm}}{\sh2}.
\label{eq:HA2}
\ee
and $\hat{s}$ is the square of the c.m. subprocess energy. The normalisation factor $N$ is given in the Appendix.

\begin{figure} [h]
\begin{center}
\includegraphics[height=6cm]{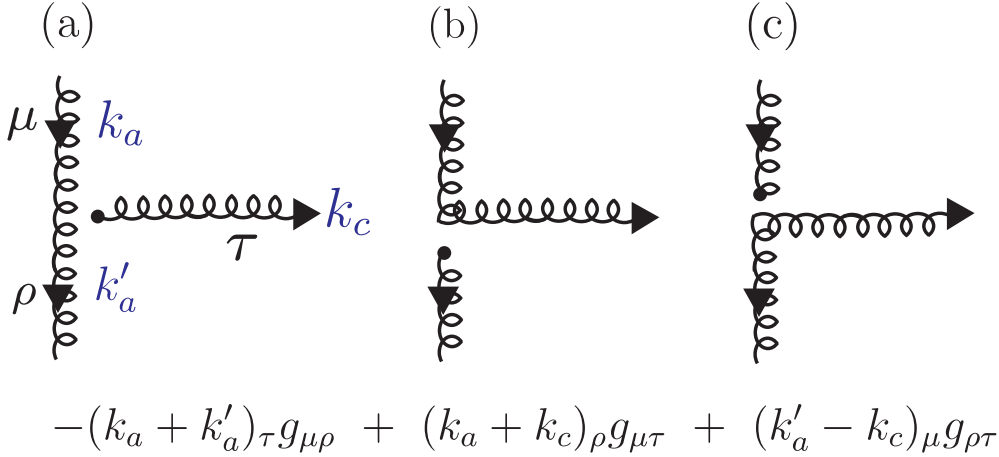}
\caption{\sf The three terms contained in the upper triple-gluon vertex, $\Gamma_{\mu\rho\tau}$, of Fig.\ref{fig:a3} where $\mu,\rho,\tau$ denote the gluon polarisations. There is a similar structure for the lower triple-gluon vertex. }
\label{fig:a4}
\end{center}
\end{figure}

As a rule, soft gluon emission does not change the polarisation structure of the amplitude, and a pseudoscalar particle should not be created in the collision of two gluons in a P-even state.  So why is pseudoscalar $\phi_-$ production in $pp \to p+ g\phi g+p$ possible?
To gain insight into why occurs, it is informative to look at the structure of the two triple-gluon vertices in Fig.\ref{fig:a3}. 
Recall that a triple-gluon vertex which is responsible for gluon emission contains three terms, as shown in  Fig.\ref{fig:a4}.

Here, diagram (a) preserves the helicity, $\lambda_{a'}=\lambda_a$, whereas the sum of the last two diagrams flips the helicity\footnote{For instance in diagram (b) of Fig.\ref{fig:a4} the incoming polarisation vector ${\vec \epsilon}_a$ goes to gluon $c$, while gluon $a'$ gets a polarisation ${\vec \epsilon}_{a'}$ in the direction of ${\vec k}_{ct}$. Since ${\vec \epsilon}_c$ is orthogonal to ${\vec k}_{ct}$, this means that ${\vec \epsilon}_a$ is orthogonal to ${\vec \epsilon}_{a'}$. Similarly for diagram (c). From the sum (b)$+$(c) we find  $\lambda_{a'}=-\lambda_a$.}, $\lambda_{a'}=-\lambda_a$. However, the contribution of (a) is enhanced relative to (b)+(c) by the factor\footnote{This is a well-known result in the classic Weis\"{a}cker-Williams approach \cite{WW},
where the polarisation vector of a gluon can be replaced by $k_t/x$. The $1/x$ enhancement can be checked easily using the explicit form of the triple-gluon vertex displayed in Fig.\ref{fig:a4}, and the fact that the polarisation vectors of the gluons are orthogonal to their momenta.} $1/x_c^+$, where $x_c^+$ is the light-cone fraction of the momentum $k_a$ of the incoming gluon carried by gluon $c$. Recall that, according to the $J_z=0$, P-even selection rule, the incoming gluons $a$ and $b$ have the same helicity, $\lambda_a=\lambda_b$. This configuration corresponds to the production of the scalar boson $\phi_+$, whereas we need $\lambda_{a'} \ne \lambda_{b'}$ in order to saturate the Levi-Civita tensor to produce a pseudoscalar $\phi_-$. Thus, the production of a scalar $\phi_+$ is driven by diagram (a) of the triple-gluon vertex, whereas production of a pseudoscalar $\phi_-$ comes from diagrams (b)+(c) in one of the two three-gluon vertices. Since (b)+(c) is suppressed by the factor $x_c^+$ (or $x_d^-$ if it is the lower gluon vertex which provides the pseudoscalar character) the amplitude $\cal{M}(\phi_-)$ vanishes as $\hat{s} \to m^2_{\phi_-}$, see (\ref{eq:HA2}) or the derivation in the Appendix.

\section{$A$ and $h,~H$ production in semi-inclusive processes}

As an example, we consider scalar and pseudoscalar neutral Higgs boson production in the minimal supersymmetric model (MSSM), which contains three neutral Higgs bosons: the light $h$ and heavy $H$ scalars, and the pseudoscalar $A$, \cite{reviews}.
In a large area of tan$\beta-m_A$ parameter space, the pseudoscalar boson $A$ is practically degenerate with
one of 
the scalar bosons, not only in mass and width, but also in the branching ratios of the main decay channels. Therefore, it will be challenging to separate $A$ from $H$ (or $h$), and to know what we have observed.

For the purely exclusive process, $pp \to p+\phi +p$, the ``$J_z=0$, P-even'' selection rule effectively filters out the $A$ boson production, leaving just the $h$ and $H$ signal. However, in the semi-inclusive process, $pp \to p+g\phi g+p$, the pseudoscalar is much less suppressed, see \cite{KMRextend}.
In comparison with exclusive scalar Higgs production, $gg^{PP} \to H$, cross section of the semi-inclusive subprocess, $gg^{PP} \to gHg$, contains the QCD coupling factor $\alpha_s^2$, but it is enhanced by the logarithmically large phase space, available for the final gluons and by a larger semi-inclusive luminosity. The latter occurs because the suppression caused by the Sudakov-like $T$ factor is much weaker, since we can permit radiation of additional softer gluons, $gg^{PP} \to gHg+n g_{\rm soft}$. The extra soft emission does not essentially change either the kinematics or the polarisation structure of the process. Thus, for scalar $H$ production we expect a cross section larger than that for pure CEP process. In addition, recall that for the allowed values of tan$\beta$, the coupling of Higgs-like bosons to gluons is larger than that in the SM.

Now what region of phase space will maximise the cross section, (\ref{eq:sighat}), for semi-inclusive pseudoscalar $A$ production? We take the kinematics where one gluon, say gluon $c$, is soft, and contains both the collinear $(dk^2_{ct}/k^2_{ct})$ and the soft $(dx_c/x_c)$ logarithms. In particular, in the interval $3<k_{ct}<25$ GeV we will have ln$k^2_{ct} \sim 4$, and ln$x_c \sim 2$. This is enough to compensate the small $\alpha_s/\pi$ factor. For the second gluon, gluon $d$, we may take a similar choice for the collinear log, but we need a relatively large $x_d$ to make $\hat{s}$ sufficiently large to avoid too much suppression from the factor $(1-m_A^4/\sh2)=(2x_d-x_d^2)$ of (\ref{eq:HA2}). This means that in the case of $A$ production we should observe an energetic jet carrying about  a half of
Pomeron-Pomeron collision energy.

Consider a relatively light pseudoscalar $A$ of mass 115 GeV, which would be almost degenerate with the light scalar $h$. If we take the default parameters of FeynHiggs \cite{feynhiggs} including $\mu$=200GeV, and tan$\beta\sim$30, then we find $\Gamma(A\to gg)$=0.9 MeV. With standard CMS/ATLAS kinematics at the LHC, the integrated semi-inclusive luminosity\footnote{Here we have used the perturbative formalism of Refs. \cite{KMRpros,KMRextend} to calculate the luminosity. The luminosity generated by double-Pomeron-exchange (DPE), where two soft Pomerons produce the active gluons according to the known diffractive PDFs, is more than two orders of magnitude lower; so it is neglected here.} is
\be
\frac{d\cal{L}}{dy}~\Delta y~=~10^{-2}.
\ee
That is, we will have one $gg^{PP}$ pair of active gluons for every 100 $pp$ collisions at 14 TeV.

The effective (observable) cross section of semi-inclusive $A$ production, $\sigma_{\rm obs}=\cal{L}\hat{\sigma}$, is about
\be
d\sigma_{\rm obs}~=~12{\rm fb}\cdot \frac{\alpha_s}{\pi}~\frac{dk^2_{dt}}{k^2_{dt}}~\frac{dx_d}{x_d}~|{\cal M}(A)|^2,
\ee
where $\cal{M}(A)$ is given in (\ref{eq:HA2}). Here we have not integrated over the phase space of the hardest gluon ($d$) emission, but have included the integration over the softer emitted gluon ($c$) in the 12fb. The kinematical factor $1/(1-x^-_d)$ in (\ref{eq:xc}) is essentially cancelled by the approximate  $1/\hat{s}$ dependence of $\cal{L}$ in this kinematic domain. If we integrate over the relevant phase space of the hard gluon, then we obtain $\sigma_{\rm obs}(A)\sim$ 4fb. Unfortunately for semi-inclusive $A \to b\bar{b}$ production and decay, the QCD $b\bar{b}$ background is not suppressed by the $J_z=0$ selection rule (unlike pure exclusive production). Therefore we resort to the $A \to \tau\tau$ decay mode, which has a branching fraction of about 10$\%$. The final semi-inclusive cross section,
\be
 \sigma_{\rm obs}(A \to \tau\tau)\sim 0.4{\rm fb},
\ee
 is small, but not hopeless for a large LHC luminosity\footnote{We can expect that dedicated fast-timing proton detectors with a
few pico-second  resolution (see \cite{FP420}),
and  additional experimental cuts, 
 will help to overcome pile-up problems,
see \cite{bonnet,KP2}, \cite{CLP}$-$\cite{tripl}. Moreover,
there are ideas that at the next stage(s) of the CMS upgrade, it may be possible to double the trigger
latency of the experiment, which  would  allow for the design of a level-1
trigger for the proton spectrometers at 420 m \cite{latency}.}. 

The main background is the semi-inclusive process
$Z \to \tau\tau$ + 2 quark jets, arising from 
the Double-Pomeron-Exchange (DPE), with two
quarks in the hard matrix element $q\bar q\to Z$ coming from the PDFs of the soft Pomeron. Using the known diffractive PDFs,
we find that the corresponding cross section ($\sim 0.5$fb) is comparable to that of the $A
\to \tau\tau$
signal. This background contribution can be calibrated using the
$Z \to \mu\mu$ and  $Z \to ee$ modes.
Note that in the standard ATLAS inclusive searches \cite {atl}, for MSSM Higgs bosons
in the mass range $110-130$ GeV,  the dominant background is also
$Z \to \tau \tau$. Although the measured $m_{\tau \tau}$ mass distribution is broad, a mass
window $111<m_{\tau \tau}<198$ GeV can be imposed in order to reduce the
background by about a factor of 5 \cite{atl}.
A similar procedure could be applied in our semi-inclusive case, with the $\tau\tau$ signal extracted in a similar way to the inclusive search\footnote{We are grateful to
Andy Pilkington for discussions of these issues.}.
With such a favourable expected
signal-to-background ratio, $S/B$, it will be important to determine
the experimental strategy, such as triggering, event selection
and cuts, which, in particular, will allow a strong reduction of the overlap background
caused by pile-up events. It is worth mentioning that, unlike inclusive MSSM Higgs searches in the $\tau\tau$ mode, in our case the $t{\bar t}$ background is very small, and can be neglected.

For a heavier pseudoscalar $A$ boson of mass 140 GeV, where now $A$ and $H$ are approximately degenerate, $\sigma_{\rm obs}$ is about twice smaller, using the same SUSY parameters. However, in some exotic SUSY scenarios (such as $\mu=-$700 GeV, for example, \cite{HKRSTW})
the semi-inclusive $A$ production cross section can be 10 times larger.

The presence of the  pseudoscalar $A$ boson signal may be revealed by the threshold behaviour. If we neglect boson width effects, then we expect just after threshold for scalar $H$ production that the cross section will decrease due to the factor $|{\cal M}(H)|^2$ of (\ref{eq:HA2}), while the existence of a pseudoscalar $A$ will make the decrease much slower. Indeed, in the case of degenerate $A$ and $H$ bosons, we have
\be
\sigma_H~:~\sigma_A~:~\sigma_{A+H}~~~\simeq~~~(1-\varepsilon)^2~:~\varepsilon^2~:~1-2\varepsilon+2\varepsilon^2,
\ee
where $\varepsilon=x-x^2/2$ and $(1-x)=m^2/\hat{s}$, with the threshold at $x=0$.

\section{Identification of low mass Goldstone bosons}

A light pseudoscalar Goldstone boson, $a$, with a mass in the range of a few GeV, is not excluded by experiment\footnote{An upper mass limit, 2$m_B$, is usually required in order
to avoid the constraints from LEP on the light Higgs  decay $h\to 2a \to 4b$.
Lower mass limits come from the searches for new light
CP-odd particles  in radiative $\Upsilon(n{\rm S})\to\gamma a$ decays (see, for
example, \cite{cleo})
or from the searches for $h\to 2a$ decay chains at the Tevatron (see, for
example, \cite {Tev}).
Note, that as discussed in \cite{cc2}, the coupling of the pseudoscalar $a$
to down-type quarks and charged leptons can
be strongly  suppressed. In such a case,
the upper limit for the mass
of the pseudoscalar $a$ could be as high as $M_h$/2.},
because it couples very weakly to light
flavour fermions,
see, for example \cite{dg}~-~\cite{cc2}.
In some extensions of the MSSM the dominant
decay of the lightest Higgs boson is $H \to aa$, with $a \to \tau\tau$ \cite{dg} or to
$c\bar{c}$ \cite{cc1}, or to $gg$ \cite{cc2}. This, in particular, will allow
a reduction in tension  between the LEP limit on
the Higgs mass and fine tuning, which 
is needed in order for the LEP Higgs mass limit
to be compatible with MSSM. A
light pseudo-Goldstone boson, that couples significantly
to the Higgs boson, is required also in many other theories, such as
little Higgs models and non-supersymmetric two-Higgs-doublet models.
Light pseudoscalars $a$ are also suggested by superstring modelling.

For example, the global symmetry breaking scenario, discussed
in \cite{cc1}, implies the existence of a few-GeV pseudo-Goldstone boson $a$, with a dominant
 decay into two gluon jets, $a \to gg$. Now the $a$ bosons that are produced in $H$ decay, may have  relatively large transverse momenta, $p_{at}\sim40$ GeV.  Thus, the opening angle between the two gluon jets in an $a \to gg$ decay, $\theta \sim 2m_a/p_{at}\sim 0.4$, is smaller than the size of the jet cone usually used in the jet searching algorithm. In this case, the system, $H \to aa \to (gg)(gg)$, will be seen experimentally as two broad jets, each composed of a gluon pair. The unusual feature of these gluon pairs is that each pair forms a colour singlet. Actually a gluon pair is a short, light $(m_{\rm string}\simeq m_a)$ {\it colour string}, boosted in the transverse direction. The multiplicity of this system is much lower than for the usual gluon jet with the same $p_t$. After the Lorentz boost, we have almost no low $p_t$ particles. Therefore, it would be interesting to search for events with such special characteristics in semi-inclusive processes, where the background of low $p_t$ hadrons, arising from multiple interactions, is absent. Moreover, the $a \to gg$ jet may be revealed as a narrow peak, $m_{\rm jet}\simeq m_a$ in the distribution over the jet mass, measured in a way similar
to that proposed in Ref. \cite{BES}.

\section{Conclusions}

Here, we have discussed the advantages of semi-inclusive processes with two fast tagged forward outgoing protons, in the search for new BSM phenomena. The limited energy transfer to the central system allows the observation of all the secondaries in the central detector, which, in this case, therefore, covers practically the whole $4\pi$ domain. Despite the low cross sections, this offers an attractive tool to observe (a) events with missing longitudinal 4-momentum (as in invisible Higgs production), and (b) new BSM effects (in particular, the production of possible pseudoscalar Goldstone bosons $\phi$) in Pomeron-Pomeron collisions selected by semi-inclusive kinematics, $pp \to p+(g\phi n_{\rm soft})+p$. The $+$ signs denote large rapidity gaps, and $n_{\rm soft}$ indicates any number of soft particles (in the central detector). An advantage over inclusive production is that there is a much lower background from underlying events.

Since such BSM particle searches may be done using the
 planned forward proton detectors at the LHC, we believe this opportunity to look for new neutral heavy objects should be 
given a serious consideration.

\section*{Appendix}
The tree-level amplitudes of the subprocess $gg \to g \phi g$ may be evaluated in the helicity formalism \cite{MHV}
\footnote{We thank Ciaran Williams for discussion of MHV rules.}.
They take the form
\be
M(\phi)~=~Cg^2\sum_{\rm perm} ~{\rm Tr}(t^at^bt^ct^d)~m(\phi,k_a^{\lambda_a},k_b^{\lambda_b},k_c^{\lambda_c},k_d^{\lambda_d}),
\ee
where the constant $C$ is the coupling of the effective $gg \to \phi$ vertex (see (\ref{eq:ggV})); $t^a$ are colour matrices of the fundamental representation of SU(3); $g$ is the QCD coupling; and $\lambda^a$ and $k_a$ are the helicities and momenta of the corresponding gluons. If we use the Ward identity
\be
m(\phi,a,b,c,d)~+~m(\phi,b,a,c,d)~+~m(\phi,b,c,a,d)~=~0,
\ee
and the fact that the two incoming gluons, $a$ and $b$, form a colour singlet, then the final result may be expressed in terms of just one of these three amplitude forms -- $m(\phi,b,c,a,d)$. We are concerned with the case when the initial state has $J_z=0,$ P-even. That is, when both of the incoming gluons have the same helicity. Then only two helicity amplitudes contribute \cite{MHVH}
\be
m(\phi,-,-,-,-)~=~\frac{m^4_\phi}{\sqrt{s_{ac}s_{cb}s_{bd}s_{da}}}
\label{eq:b1}
\ee
and
\be
m(\phi,+,-,+,-)~=~\frac{s^2_{ab}}{\sqrt{s_{ac}s_{cb}s_{bd}s_{da}}},
\label{eq:b2}
\ee
where $s_{ij}=(k_i+k_j)^2$ denote the energy squared of the $ij$ gluon pair. As we are looking for a logarithmically ($\int dx_c/x_c$) enhanced cross section, and we are assuming that at least one of secondary gluons
is soft, we have neglected a second contribution to (\ref{eq:b2}),  which is suppressed by the factor $(x_c x_d)^2 \ll 1$. Also we neglect the possibly complex phases of the amplitudes $m$, since they are the same for both amplitudes.

The matrix elements for $\phi_{\pm}$ production are given by the sum (difference) of the amplitudes (\ref{eq:b1}) and  (\ref{eq:b2}). We choose the $z$-axis in the direction of the incoming gluons in the c.m. frame. The cross section of the hard $gg \to g_c\phi g_d$ subprocess is then
\be
d{\hat \sigma}_{\phi}~=~N~\frac{dk_{ct}^2}{k^2_{ct}}~\frac{dx_c}{x_c}~\frac{dk_{dt}^2}{k^2_{dt}}~\frac{dx_d}{x_d}~|{\cal M}(\phi)|^2
\ee
where
\be
{\cal M}(\phi_\pm)~=~1\pm\frac{m^4_{\phi_\pm}}{\sh2}.
\label{eq:HA}
\ee
The normalisation factor $N$ includes the QCD coupling, $\alpha_s$, and the $gg \to \phi$ vertex factor $C$ of (\ref{eq:ggV}). It can be expressed in terms\footnote{Compare eq.(34) of \cite{KMRpros}.} of the $\phi \to gg$ decay width $\Gamma_{gg}$:
\be
N~=~\alpha^2_s\frac{N_C^2}{2(N^2_C-1)}\frac{\Gamma_{gg}}{m^3_\phi}~\delta \left(1-\frac{(1-x_c^+)(1-x_d^-)\hat{s}}{m^2_\phi}\right).
\ee
Here $\hat{s}=s_{ab}=(k_a+k_b)^2$ is the energy squared of the hard subprocess; $\vec{k}_{tc}$ is the transverse momentum of the outgoing gluon $c$ and $x_c^+$ its light-cone momentum fraction --
\be
k_c~=~x_c^+ k_a+k_{ct}+x_c^- k_b,
\ee
and similarly for the outgoing gluon $d$
\be
k_d~=~x_d^- k_b+k_{dt}+x_d^+ k_a.
\ee
We assume that gluons $c$ and $d$ are emitted in the hemispheres of the incoming gluons $a$ and $b$ respectively; that is $x_c^-$ and $x_d^+$ are small $(x_c^- < x_c^+ , ~x_d^+ < x_d^-)$. 

On convoluting with the effective Pomeron-Pomeron luminosity, ${\cal L}(gg^{PP})/d{\rm ln}\hat{s}$, we have to integrate over $\hat{s}$.
 We, therefore, arrive at the form
\be
d{\hat \sigma}_{\phi}~=~\frac{dk_{ct}^2}{k^2_{ct}}~\frac{dx_c}{x_c}~\frac{dk_{dt}^2}{k^2_{dt}}~\frac{dx_d}{x_d}~\frac{\alpha^2_s}{(1-x_c^+)(1-x_d^-)}~\frac{N_C^2}{2(N^2_C-1)}~\frac{\Gamma_{gg}}{m^3_\phi}~|{\cal M}(\phi)|^2
\label{eq:xc}
\ee

\section*{Acknowledgements}

We thank Albert De Roeck, Csaba Csaki, Risto Orava, Krzysztof Piotrzkowski
 and Ciaran Williams, and especially Andy Pilkington, for valuable discussions.
MGR would like to thank the IPPP at the University of Durham for hospitality. This work was supported by the grant RFBR
10-02-00040-a, by the Federal Program of the Russian State RSGSS-3628.2008.2.

\end{document}